\documentclass[sn-mathphys]{sn-jnl}

\usepackage{subfigure}

\jyear{2021}%

\theoremstyle{thmstyleone}%
\theoremstyle{thmstyletwo}%

\theoremstyle{thmstylethree}%

\floatname{algorithm}{Processing Flowchart}

\begin{document}

\title[Article Title]{High performance reconciliation for practical quantum key distribution systems}


\author[1]{\fnm{Hao-Kun} \sur{Mao}}\email{mao\_hit@qq.com}
\author*[1]{\fnm{Qiong} \sur{Li}}\email{qiongli@hit.edu.cn}
\author[2,3]{\fnm{Peng-Lei} \sur{Hao}}\email{haopl@mail.ustc.edu.cn}
\author[4]{\fnm{Bassem} \sur{Abd-El-Atty}}\email{bassimeldeeb@gmail.com}
\author[5,6,7]{\fnm{Abdullah M.} \sur{Iliyasu}}\email{a.iliyasu@psau.edu.sa}

\affil*[1]{\orgdiv{Department of Computer Science and Technology}, \orgname{Harbin Institute of Technology}, \orgaddress{\city{Harbin}, \postcode{150080}, \country{China}}}
\affil[2]{\orgdiv{CAS Key Laboratory of Quantum Information}, \orgname{University of Science and Technology of China}, \orgaddress{\city{Hefei}, \postcode{230026}, \country{China}}}
\affil[3]{\orgdiv{CAS Center For Excellence in Quantum Information and Quantum Physics}, \orgname{University of Science and Technology of China}, \orgaddress{\city{Hefei}, \postcode{230026}, \country{China}}}
\affil[4]{\orgdiv{Department of Computer Science, Faculty of Computers and Information}, \orgname{Luxor University}, \orgaddress{\country{Egypt}}}
\affil[5]{\orgdiv{School of Computing}, \orgname{Tokyo Institute of Technology}, \orgaddress{\city{Yokohama}, \postcode{226-8502}, \country{Japan}}}
\affil[6]{\orgdiv{Electrical Engineering Department, College of Engineering}, \orgname{Prince Sattam Bin Abdulaziz University}, \orgaddress{\city{Al-Kharj}, \postcode{Al-Kharj}, \country{Saudi Arabia}}}
\affil[7]{\orgdiv{School of Computer Science and Technology}, \orgname{Changchun University of Science and Technology}, \orgaddress{\city{Changchun}, \postcode{130022}, \country{China}}}


\abstract{Quantum key distribution (QKD) is a promising technique for secure communication based on quantum mechanical principles. To improve the secure key rate of a QKD system, most studies on reconciliation primarily focused on improving the efficiency. With the increasing performance of QKD systems, the research priority has shifted to the improvement of both throughput and efficiency. In this paper, we propose a high performance solution of Cascade reconciliation, including a high-throughput-oriented framework and an integrated-optimization-oriented scheme. Benefiting from the fully utilizing computation and storage resources, effectively dealing with communication delays, the integrated-optimization-oriented parameters setting, etc., an excellent overall performance was achieved. Experimental results showed that, the throughput of up to 570Mbps with an efficiency of 1.038 was achieved, which, to our knowledge, was more than four times faster than any throughput previously demonstrated. Furthermore, throughputs on real data sets were capable of reaching up to 86Mbps even on embedded platforms. Additionally, our solution offers good adaptability to the fluctuating communication delay and quantum bit error rate (QBER). Based on our study, low performance (i.e. low power-consumption and cost-effective) CPU platforms will be sufficient for reconciliation in the existing and near-term QKD systems.}




\maketitle

\section{Introduction}
\label{intro}
Quantum key distribution (QKD) is a promising technique for distributing unconditionally secure keys between remote parties in real time \cite{1_Bennett_2014}. Although QKD systems can theoretically contribute towards enhancing the security of the communication systems, their practical applications are constricted due to their low secure key rates and high costs \cite{2_Yuan_2018, 3_Duplinskiy_2018}. To address this issue, most research focused on the optimizations of the two major QKD layers, i.e., the so called photonic layer and post-processing layer \cite{2_Yuan_2018}. The photonic layer has in the past been considered the biggest impediment to improving the secure key rate. However, with the recent advances in single-photon detector technologies \cite{4_Boaron_2018}, photonic integrated circuits \cite{5_Pirandola_2020} and other key technologies \cite{6_Lucamarini_2018, 7_Yin_2016}, the performance bottleneck is gradually shifting to the post-processing layer \cite{2_Yuan_2018, 8_Zhang_2020}. Thus, as the major performance-limiting module in the post-processing layer,  reconciliation has attracted extensive attentions \cite{9_Dixon_2014, 10_Wang_2018, 11_Gao_2019, 12_abd2020controlled}. In practice, the reconciliation module has two main performance metrics: efficiency (i.e. the ratio of actual transmitted information to the necessary amount of information) and throughput (i.e. the amount of data that can be processed per second) \cite{13_Pedersen_2015}. In the early stage of development, most reconciliation studies focused solely on efficiency because the efficiency is the dominant factor for the secure key rate and the requirement for throughput is easy to meet due to the limited performance of the photonic layer. With the increasing performance of the photonic layer, the requirement for the throughput of reconciliation increases sequentially. Therefore, in high-speed QKD systems, not only efficiency but also throughput should be taken into consideration \cite{2_Yuan_2018, 14_abd2020quantum, 9_Dixon_2014}. Unfortunately, the above two metrics are often inversely correlated \cite{15_abd2019efficient, 16_abd2020secure}, which makes it challenging to simultaneously improve throughput and efficiency. Consequently, this study focusses on the integrated optimization of efficiency and throughput of the reconciliation module in discrete-variable (DV) QKD systems which are more widely deployed.

In general, the reconciliation protocols in DV-QKD systems can be divided into two categories: interactive and non-interactive \cite{17_Li_2014}. The most widely used interactive reconciliation protocol was Cascade \cite{18_Brassard_1993}. In this protocol, the correlated sequences belonging to the two remote parties were first shuffled and divided into blocks. Then, the errors were detected and corrected by comparing the parity bits of these blocks and performing binary search operations, respectively. The aforementioned operations had to be performed iteratively for a number of passes to correct further errors. In addition, the Cascade protocol utilized the relationships among passes to correct more errors by involving backtracking operations. Although a number of communication rounds were required in the Cascade protocol, the advantages of simplicity and relatively high efficiency were obvious \cite{20_Mao_2019}. For further improvement of efficiency, some later efforts concentrated on the modifications and optimizations of the original Cascade protocol \cite{21_Jesus_2015, 22_Yan_2008, 23_Pacher_2015}. However, nearly all these protocols suffered from high communication overhead, potentially limiting their applications in practical QKD systems \cite{9_Dixon_2014}. Accordingly, the non-interactive reconciliation protocols based on forward error correction methods, such as low density parity check (LDPC) \cite{25_David_2011, 26_Jesus_2012, 27_Kiktenko_2017,  29_Li_2020, 30_Yang_2020} or polar codes \cite{31_Paul_2014, 32_Yan_2018, 33_Kiktenko_2020}, were proposed. Due to the much less of communication rounds, the non-interactive protocols showed good performance even when the communication delay was high.

Although research of non-interactive protocols have received much attention, this study is devoted to the high-performance Cascade reconciliation based on the following considerations. On one hand, the high communication overhead, which was considered to be the main drawback of Cascade protocols, has relatively low impact on the reconciliation performance required by most realistic deployments of QKD. For instance, in a most common fiber-based QKD system, a low latency communication channel is available by either multiplexing quantum and classical channels, or just by using a different fiber in the same fiber bundle. Therefore, the total end-to-end latency is close to only about 1ms \cite{13_Pedersen_2015}. Under such a realistic condition, an implementation of the original Cascade protocol on two Intel i7 processors achieved a throughput of up to 82Mbps with an efficiency of 1.125 \cite{13_Pedersen_2015}. The above results validated the potential of Cascade protocols to reach high throughput despite the adverse effects caused by high communication overhead. Moreover, the increase in authentication cost of a practical QKD system is usually negligible, even though multiple communication rounds are required in Cascade reconciliation. This is because the remote parties do not need to check the authenticity of communications over the classical channel per round. Instead, the authenticity is usually checked by an exchange of hash values of the whole incoming traffic after the privacy amplification (PA) stage \cite{3_Duplinskiy_2018, 34_Fung_2010}. Meanwhile, on the other end, in addition to the inherent advantages of Cascade, such as simplicity and high efficiency,  Cascade also has several other advantages over non-interactive protocols. First, the rateless feature of Cascade \cite{21_Jesus_2015} results in better adaption to the QBER variations. Second, the QBER range, within which Cascade protocols perform well, is just the focus of most DV-QKD systems \cite{5_Pirandola_2020}. Third, the characteristic of low computation complexity makes Cascade well suited to work together with a PA module which requires significant computation resources \cite{35_Yan_2020}. Therefore, put all together, we believe that Cascade reconciliation will be a better choice for most real-world scenarios than non-interactive protocols.

Consequently, this study is primarily aimed at improving both the throughput and efficiency via a high performance solution for Cascade reconciliation. In this regard, the main contributions of our work are as follows. First, a high-throughput-oriented framework is proposed to increase utilization of computation resources and effectively deal with communication delay. Second, on the basis of the framework, we propose an integrated-optimization-oriented scheme, including the optimizations of core parameters, storage scheme and backtracking process. Results of experiments reported show that a throughput of 570Mbps with efficiency of 1.038 is achieved at 1\% QBER by using two i7-6700HQ CPUs connected by a 50 km long optical fiber. This is more than four times the throughput of the fastest reconciliation protocol known to us.

The rest of this paper is organized as follows. Section \ref{related} introduces several measures for reconciliation efficiency and a brief description of a highly efficient improved Cascade protocol available in the literature. Section \ref{sec:3} presents the details of our proposed high performance solution, while its throughput and efficiency results are reported and analyzed in Section \ref{experiment}.

\section{Reconciliation Efficiency}
\label{related}
Assuming random variables $A$ and $B$ represent Alice's and Bob's sequences of length $n$, respectively. Then according to the noiseless coding theorem, the minimum amount of exchanged information for successful reconciliation is given by the conditional entropy $H\left(A \mid B\right)$ \cite{21_Jesus_2015}. In a DV-QKD system, the discrepancies between $A$ and $B$ can be assumed to the consequence of a transmission over a binary systematic channel (BSC) with crossover probability $\varepsilon$  which is usually referred to as QBER. Therefore,  $H\left(A \mid B\right)$ can be written as $nh(\varepsilon)$, where the binary Shannon entropy $h(\varepsilon ) =  - \varepsilon {\log _2}(\varepsilon ) - (1 - \varepsilon ){\log _2}(1 - \varepsilon )$. Let $m$ be the length of the leaked information, the reconciliation efficiency $f$ can be defined as:

\begin{equation}
\label{eq:1}
f = \frac{m}{{nh(\varepsilon )}}
\end{equation}

Since $nh(\varepsilon)$  is the minimum length of exchanged information, a smaller $f$ means a better reconciliation and $f=1$ stands for the perfect reconciliation. Let  $R$ denotes the ratio of transmitted information to the code length, such that $R = 1 - m/n$. Then, the efficiency $f$ can be rewritten as:

\begin{equation}
\label{eq:2}
f = \frac{1-R}{{h(\varepsilon )}}
\end{equation}

Following the information theory, one transmitted bit can reveal at most one bit information. Therefore, in Cascade reconciliation, the norm is to use the number of transmitted bits in one direction ${m^*}$  as an upper bound for  $m$, since ${m^*}$  may be linearly correlated and linear combinations of bits do not leak information.

In practice, the frame error rate (FER), which is the probability that reconciliation fails for a given frame, should also be taken into consideration. For non-interactive protocols, the FER may be caused by non-convergence or converging to an inappropriate frame. In the case of Cascade protocols, some sub-blocks may contain positive numbers of undetected errors after implementing reconciliation procedure, which leads to an unsuccessful reconciliation. We note that the error frame can be detected by an additional step called verification \cite{20_Mao_2019} and have to be discarded. It is also remarkable that, in practice, higher $f$ may not imply better reconciliation due to the presence of FER. Generally, the value of FER is adjusted downwards by $10^{-3}$ to avoid its negative effect. In some Cascade studies \cite{21_Jesus_2015, 36_toyran2017new}, a modified efficiency $f_{FER}$ was used by taking into account the FER. The $f_{FER}$ can be calculated as follows:

\begin{equation}
\label{eq:3}
f_{FER} = \frac{{(1 - FER)(1 - R) + FER}}{{h(\varepsilon )}}
\end{equation}

\section{High Performance Solution of Cascade Reconciliation}
\label{sec:3}
In this section, we propose a high performance solution of Cascade reconciliation, including a high-throughput-oriented framework and an integrated-optimization-oriented scheme. The proposed implementation framework effectively inhibits the throughput degradation caused by high communication delay. In addition, the optimizations of frame processing are also considered to meet the needs of high-performance parallel implementation. On the basis of this framework, we propose an implementation scheme that optimizes core parameters, storage scheme and backtracking process.

\subsection{High-Throughput-Oriented Framework}
\label{sec:3.1}
Our framework, which is similar to the blockchain technology, is an optimized integration of the existing techniques that is timely and pivotal for practical applications. The framework is schematically illustrated in Processing Flowchart \ref{algorithm:1}. To maximize the overall throughput performance, different levels presented in the proposed framework have to attain high levels of performance. The first level, located among multiple processor cores, is where several sequences are simultaneously reconciled. The main problem of this level centers on the provision of an efficient parallel processing approach. The second level is located inside the processor core and an efficient pipeline scheduling of threads is designed to minimize adverse effects of communication delay. The third level is located inside each thread and it is primarily concerned with ensuring that, wherever possible, the functional modules are in working state rather than waiting state. Detailed implementations of the three levels are presented in the remainder of the section. We note that although our framework is designed for the specific improved Cascade protocol in \cite{23_Pacher_2015}, it can be adapted to other variations of Cascade protocols with relatively minor modifications.

\begin{algorithm}
\caption{Implementation Framework}\label{algorithm:1}
\begin{algorithmic}[1]
\State{\textcolor[RGB]{47, 47, 135}{ \textbf{$\triangle$ Level 1: Parallel Processing}}}
\For{each core \textbf{par -}}
	\State{\textcolor[RGB]{47, 47, 135}{\textbf{$\triangle$ Level 2: Multi Pipeline}}}
	\For{$stage=1 \to \rm{{STAGE}_{max}}$}
		\State{\textcolor[RGB]{47, 47, 135}{\textbf{$\triangle$ Level 3: Cascade Reconciliation}}}
		\For{$round=1 \to \rm{{ROUND}_{max}}$}
			\State{$RecvNetData();$}
			\State{$\textbf{DivideBinarySearch()};$}
			\State{$\textbf{CascadeErrorCorrection()};$}
			\State{$SendNetData();$}
		\EndFor
	\EndFor
\EndFor
\end{algorithmic}
\end{algorithm}

\subsubsection{Parallel Processing}
\label{sec:3.1.1}
In theory, the parallel implementation can significantly improve the throughput performance. However, the actual performance gain, by taking advantage of the parallel implementation, is affected by many factors. For instance, the bottleneck of parallel LDPC decoding on GPU has been demonstrated to be the slow memory accesses \cite{37_Le_Gal_2016}. To limit the impact of the mass memory latency, multi-level caches were introduced in modern CPUs. However, the memory access latency varies among different levels of caches and the sizes of faster caches tend to be smaller \cite{38_Levinthal_2009}.
In view of the characteristics of the multi-cache architecture, reducing frame length and reusing cached data are efficient strategies employed to increase the memory access efficiency in parallel processing environment.

Specifically, in our implementation, we divide the original large sequence into multiple short frames with a fixed length of 64kb based on the following considerations. On one hand, the previous simulation results have shown that a rather high efficiency can be achieved with 64kb frame length, then larger frame lengths offer no significant improvement in efficiency improvement \cite{23_Pacher_2015}. On the other hand, the latter frame can reuse the cached-data from the former frames when using fixed frame length, which decreases execution time. For instance, the time-consuming shuffling operation can be speeded up by applying a pre-stored mapping table. If all frames are of the equal length, then only a single shuffling map is required. The latter frame does not need to reload the shuffling map from slower memory, which results in an improved throughput.

\subsubsection{Multi Pipeline}
\label{sec:3.1.2}
\begin{figure}[ht]
	\centering
	\includegraphics[width=0.9\textwidth]{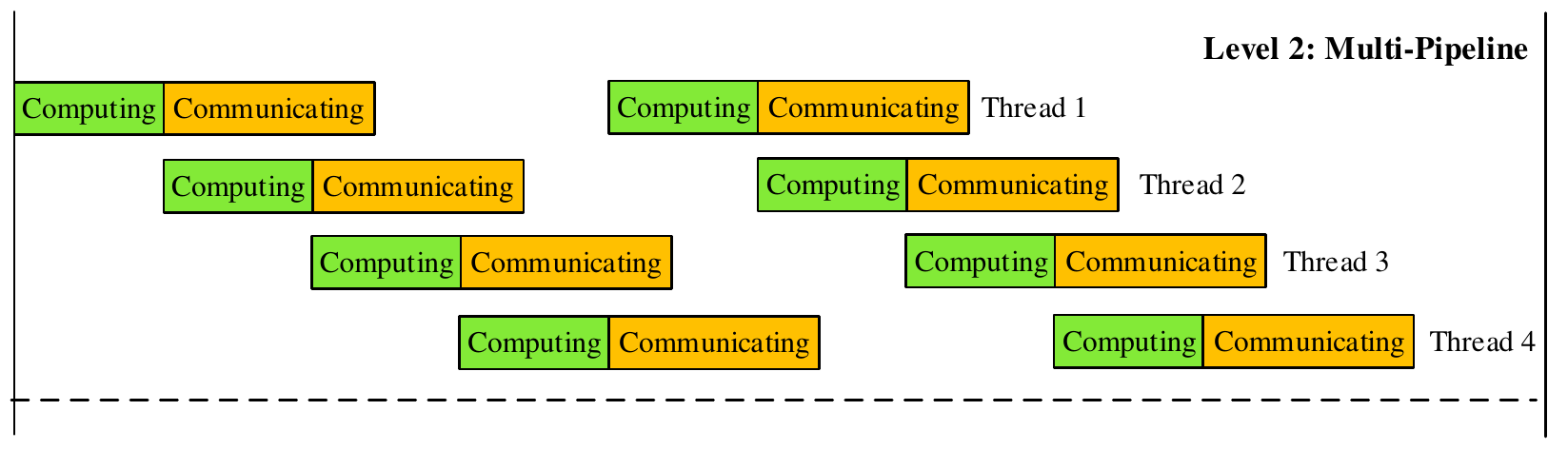}
	\caption{Layout of four-stage multi-pipeline for multi reconciliation threads}
	\label{fig:1}
\end{figure}

While its use is less commonplace in Cascade implementations, pipeline technology is one of the optimum approaches used in hardware implementations to decrease the processing delay. Moreover, it is generally held that inside each reconciliation thread, the subsequent processing is delayed pending results of the parity comparison from the other party through the commutation channel. Such a stop-and-wait scheme affects the performance and causes communication delay. To augment this, we apply multi-pipeline technology among multi reconciliation threads in this study. Fig. \ref{fig:1} presents the layout of a typical four-stage pipeline technology. The figure illustrates that applying multi-pipeline technology can help circumvent the adverse effects of the communication delay, which would result in improved throughput. Specifically, in our implementation, all the threads are executed serially to avoid the overhead of thread switching. The gains from this optimization strategy will be validated in results of experiments reported in Section \ref{experiment}.

\subsubsection{Cascade Reconciliation}
\label{sec:3.1.3}
\begin{figure}[ht]
	\centering
	\includegraphics[width=0.9\textwidth]{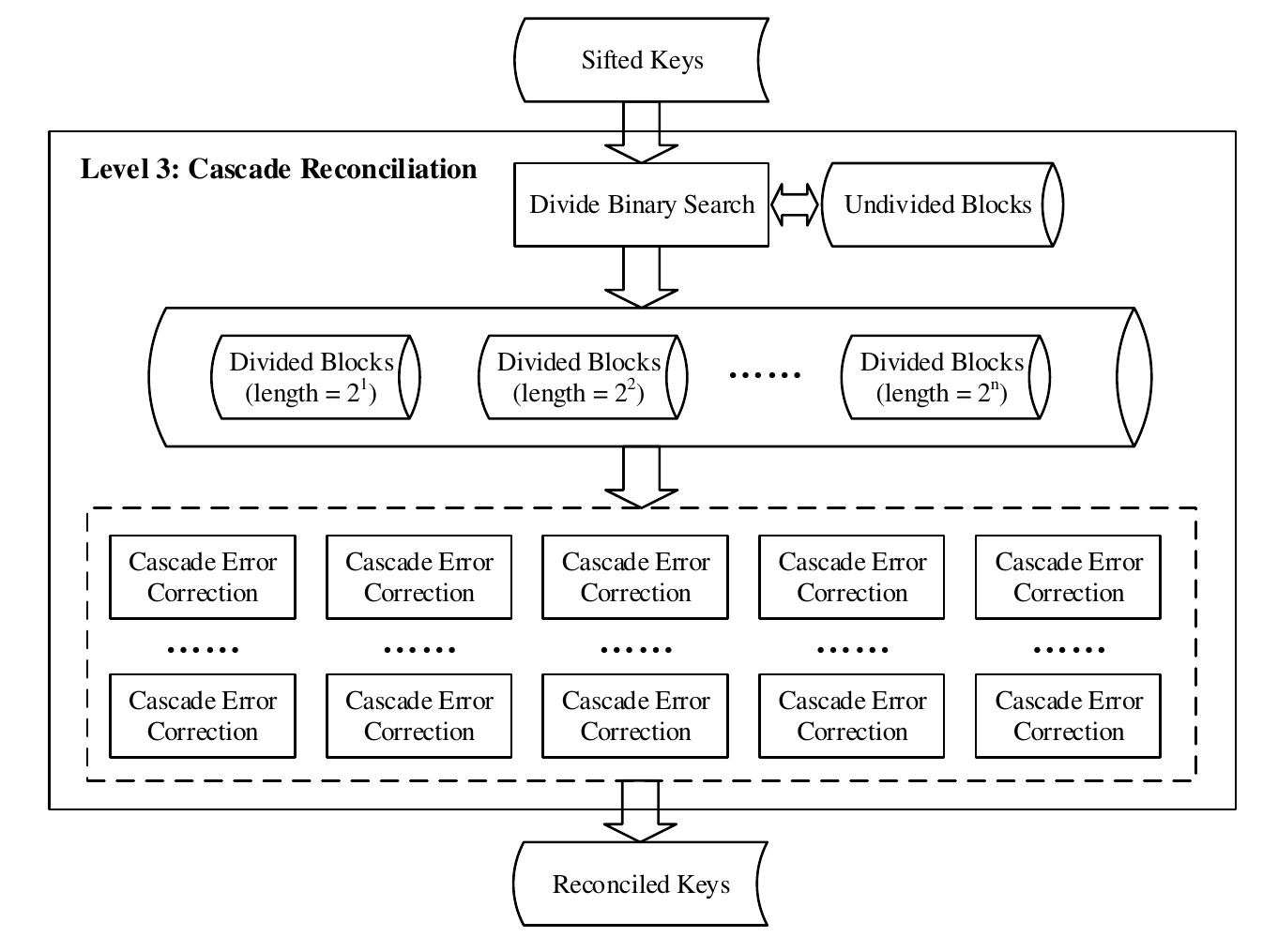}
	\caption{Flowchart for implementation of Cascade reconciliation}
	\label{fig:2}
\end{figure}

The flowchart outlining the information processing of the Cascade reconciliation module is presented in Fig. \ref{fig:2}. From it, we deduce that, upon acquisition of the data from the sifted-key buffer, the DivideBinarySearch module performs binary search operation and allocates the divided blocks into the corresponding DivideBlocks buffers. Once the size of a buffer exceeds the preset frame length of 64kb, an idle module from the CascadeErrorCorrection group will be prompted to complete the error correction task. During error correction, each module works independently by using its respective state machine and storage space, but the interactive messages of each communication round are sent and received through a uniform network packet. Finally, the CascadeErrorCorrection module will output the corrected keys into the reconciled-key buffer and quit the working state as soon as the error correction task is completed.

\subsection{Integrated-Optimization-Oriented Scheme}
\label{sec:3.2}
In this part, we mainly focus on the implementation scheme of the more complicated CascadeErrorCorrection module, including the optimizations of core parameters, storage scheme and backtracking process. 

\subsubsection{Cascade Parameter Optimization}
\label{sec:3.2.1}
Parameter optimization is long standing priority for studies on Cascade reconciliation. The original Cascade protocol calculated the block length of the first pass using the formula $0.73/QBER$, which is doubled for the subsequent three passes \cite{18_Brassard_1993}. Though the efficiency still needs improvement, the number of communication rounds was low. To further enhance the reconciliation efficiency, most modified versions of Cascade protocols used different methods to calculate the block length of the first two passes and $N/2$ for the subsequent passes. Whereas the highlighted optimized parameters were acquired with the target of improving efficiency, in this study, the optimization target is no longer the efficiency improvement but the integrated optimization of both throughput and efficiency. For this reason, the previous optimized parameters are not prioritized. Furthermore, for integrated optimization, we present an adaptive parameter setting scheme that combines the advantages of the original and modified cascade protocols. The adaptive parameter settings are presented in Table \ref{tab:1}. The block $k_{initial}$ and $k_{2,K'}$  are calculated by the Equations (\ref{eq:4}, \ref{eq:5}) as described in \cite{23_Pacher_2015}, where $K' \in \{ 1,...,\left[ {{{\log }_2}({k_{init}})} \right]\}$.

\begin{table}[ht]
\caption{Adaptive parameter setting scheme where the frame length is $N = 2^{16}$}
\label{tab:1}
\centering
\begin{tabular}{cccccc}
\hline
$k_1$ & $k_2$ & $k_3$ & $k_4$ & $k_5$ & $k_6$ \\ \hline
$k_{init}$ or $k_{init}/2$ & $k_{2,K'}$ or $k_{2,K'}/2$ & $N/16$ & $N/8$ & $N/4$ & $N/2$\\ \hline
\end{tabular}
\end{table}

\begin{equation}
 \label{eq:4}
{k_{init}} = \min ({2^{\left[ {{{\log }_2}(1/\varepsilon)} \right]}},N/2)
\end{equation}

\begin{equation}
\label{eq:5}
{k_{2,K'}} = \min ({2^{\left[ {{{\log }_2}(4/{\varepsilon_{bit}})} \right]}},N/2)
\end{equation}

\begin{equation}
\label{eq:6}
{\varepsilon_{bit}} = \varepsilon\frac{{1 - {{(1 - 2\varepsilon)}^{{2^{K'}} - 1}}}}{{1 + {{(1 - 2\varepsilon)}^{{2^{K'}}}}}}
\end{equation}

Following the layout presented, we highlight some merits of our proposed scheme from two aspects.

(1)	First two passes. The values of   $k_{init}$ and  $k_{2,k'}$  can be regarded as benchmarks for assessing efficiency since they fully account for maximizing the amount of interactive information. Therefore, to further boost adaptability of the protocol on high-performance requirements, we intend to improve the throughput performance by sacrificing the efficiency properly with smaller block lengths. To do so, we adjust  $k_1$  and $k_2$ based on $k_{init}$  and  $k_{2,k'}$ respectively, to obtain four combination modes which are presented in Table \ref{tab:2}. This approach potentially reduces the number of communication rounds, making it more adapted to high latency environments. The actual effects of the adaptive adjustment strategy will be demonstrated later in Section \ref{experiment}.

\begin{table}[ht]
\caption{Adaptive Combination of Block Lengths}
\label{tab:2}
\centering
\begin{tabular}{ccc}
\hline
Combination Mode & $k_1$ & $k_2$ \\
\hline
High-Efficiency & $k_{init}$ & $k_{2,K'}$ \\
Medium-Efficiency & $k_{init}$ & $k_{2,K'}/2$ \\
Medium-Throughput & $k_{init}/2$ & $k_{2,K'}$ \\
High-Throughput & $k_{init}/2$ & $k_{2,K'}/2$ \\
\hline
\end{tabular}
\end{table}

(2)	Subsequent four passes. For efficiency, the value of $N/2$  is commonly adopted as the block length of the subsequent passes. One substantial drawback of this approach is the low ability to detect error bits, which leads to an increasing number of passes and communication rounds. To solve this issue, in this paper, the common used $N/2$ is augmented by doubling block lengths to decrease the number of passes and communication rounds. Since the number of undetected errors in these passes is small, the negative effects on efficiency is limited but the throughput improvement is significant. 

\subsubsection{Complete-Binary-Tree-Based Storage Scheme}
\label{sec:3.2.2}
The storage consumption in Ref. \cite{13_Pedersen_2015} increases rapidly with the increasing QBER. This is attributed to the use of global backtracking list whose size grows quickly with increase in QBER. To overcome this problem, the data structure of complete-binary-tree is applied instead of the global backtracking list. In this way, most of the relevant data are stored in the same data structure and only the index information is stored in the temporary list for each pass, thereby solving the issue of rapid increase in storage consumption with QBER.

\begin{figure}[ht]
	\centering
	\includegraphics[width=0.9\textwidth]{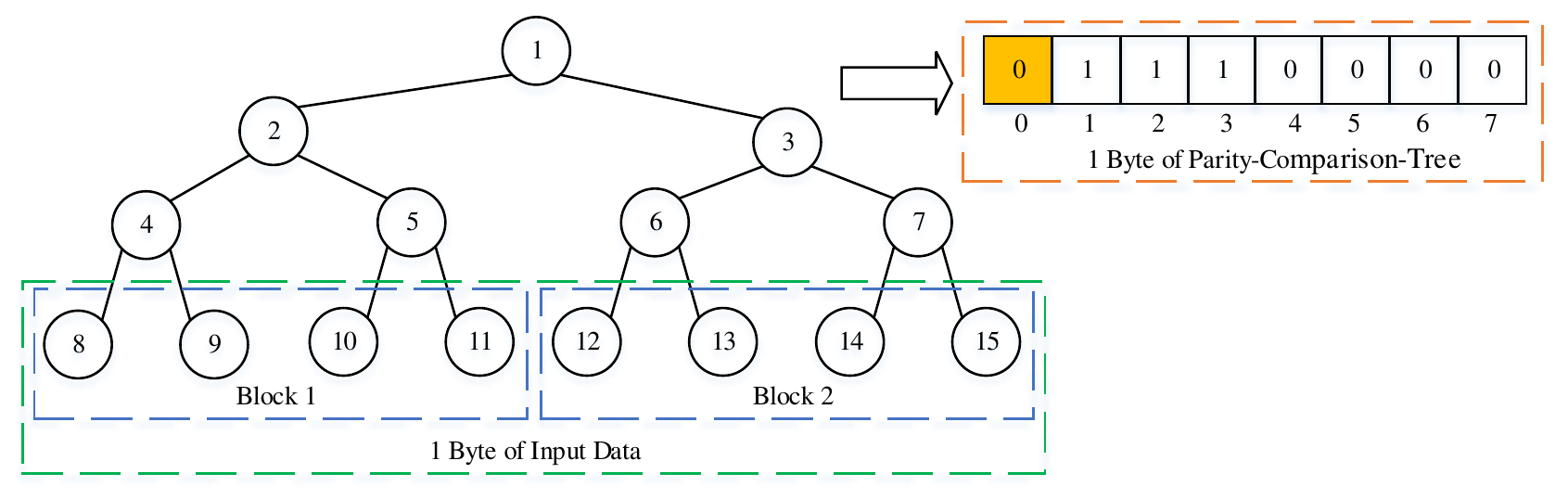}
	\caption{Schematic illustration of parity comparison tree for block length of 4}
	\label{fig:3}
\end{figure}

Specifically, to store the relevant data during error correction, we use two identical complete binary trees, which are called parity tree and parity comparison tree, respectively. Fig. \ref{fig:3} illustrates the relationships between the logical and storage structure of a parity comparison tree for a block length of 4. As demonstrated in Fig. \ref{fig:3}, the mapping of these two structures can be performed easily, which provides basis for the high-performance implementation. In terms of the storage structure, we use each byte of memory to store 8 data bits, the relevant data of the adjacent blocks are stored together. In such cases, some bits of the complete binary trees need to be marked during initialization. For instance, for block length of 4, the data bits 1, 2 and 3 are set to 1 while others are initialized to 0 as depicted in Fig. \ref{fig:3}. Since most of the errors are corrected in the first two passes, the parity comparison trees are solely in the first two passes to improve the utilization of storage resource. In this way, the total number of trees needed in the six passes decreases from 12 to 8, which is approximate a 33\% decrease.

\subsubsection{Parallel Backtracking Collision Detection}
\label{sec:3.2.3}
The backtracking collision (i.e., multiple corrected data backtrack to the same sub-block) are unavoidable, and the collisional data need to be eliminated from the backtracking list when applying parallel backtracking. This problem can be solved by querying the backtracking list. However, this approach will waste immense amounts of time which increases rapidly with QBER. Thus, the issue of how to detect backtracking collisions that need to be resolve to enhance throughput.

As presented in Fig. \ref{fig:3}, the position 0 of each binary tree is unoccupied in the memory. Based on this, we design a strategy to make reasonable use of this unused position to benefit the backtracking collision detection. The descriptions of these two trees are summarized in Table \ref{tab:3}. The parity comparison result of each complete block is stored into position 0, which ensures the backtracking procedure works normally even in the absence of the parity comparison tree. In round 3 or later, we backtrack to the complete block rather than the smallest sub-block, which offers little effect on efficiency. 

\begin{table}[ht]
\centering
\caption{Description of the complete-binary-trees}
\label{tab:3}
\resizebox{\textwidth}{1.3cm}{
\begin{tabular}{cccc}
\hline
Tree                        & Valid Passes                                                                            & Node No.                                                   & Meaning                                                                                       \\ \hline
\multirow{2}{*}{Parity Tree} & \multirow{2}{*}{All}                                                                    & 0                                                          & Parity comparison result of node 1                                                            \\ \cline{3-4}
                            &                                                                                         & \begin{tabular}[c]{@{}c@{}}Others\\ (1,2,...)\end{tabular} & Parity of the corresponding node                                                              \\ \hline
\multirow{2}{*}{Parity Comparison Tree} & \multirow{2}{*}{\begin{tabular}[c]{@{}c@{}}The previous two passes\\ (1,2)\end{tabular}} & 0                                                          & Locking identifier when backtracking                                                          \\ \cline{3-4}
                            &                                                                                         & \begin{tabular}[c]{@{}c@{}}Others\\ (1,2,...)\end{tabular} & \begin{tabular}[c]{@{}c@{}}Parity comparison result of \\ the corresponding node\end{tabular} \\ \hline
\end{tabular}
}
\end{table}

The advantage of this scheme lies in its ability to curtail the frequent and redundant querying operations that add burdens on consumption. When backtracking to each block of the first two passes, the position 0 of the parity comparison tree will be checked first. If the block is not locked, the backtracking operation will be performed directly. Otherwise, the function SearchList() will be executed to detect a collision and determine whether the data is added to or removed from the backtracking list. The actual impact of the backtracking collision detecting strategy will be demonstrated later in the results of our experiments reported in the next section.

\section{Experiments and Results}
\label{experiment}
We start the experimental validation of our proposed solution by outlining its relative realization technology. Like in \cite{20_Mao_2019, 37_Le_Gal_2016}, we apply OpenMP directives for parallel processing. For the practical implementations of shuffling and backtracking, we use an Arnold-Mapping-Based scheme \cite{40_Li_2018} whose key parameters are updated periodically. For the time-consuming operations, such as shuffling or initialization of the parity tree, the pre-calculated lookup tables are applied to decrease the processing delays. 

\subsection{Experiments for Each Optimization}
\begin{figure}[ht] 
	\centering  
	\subfigure[Throughput improvement of the multi-pipeline optimization scheme]{
		\includegraphics[width=0.41\textwidth]{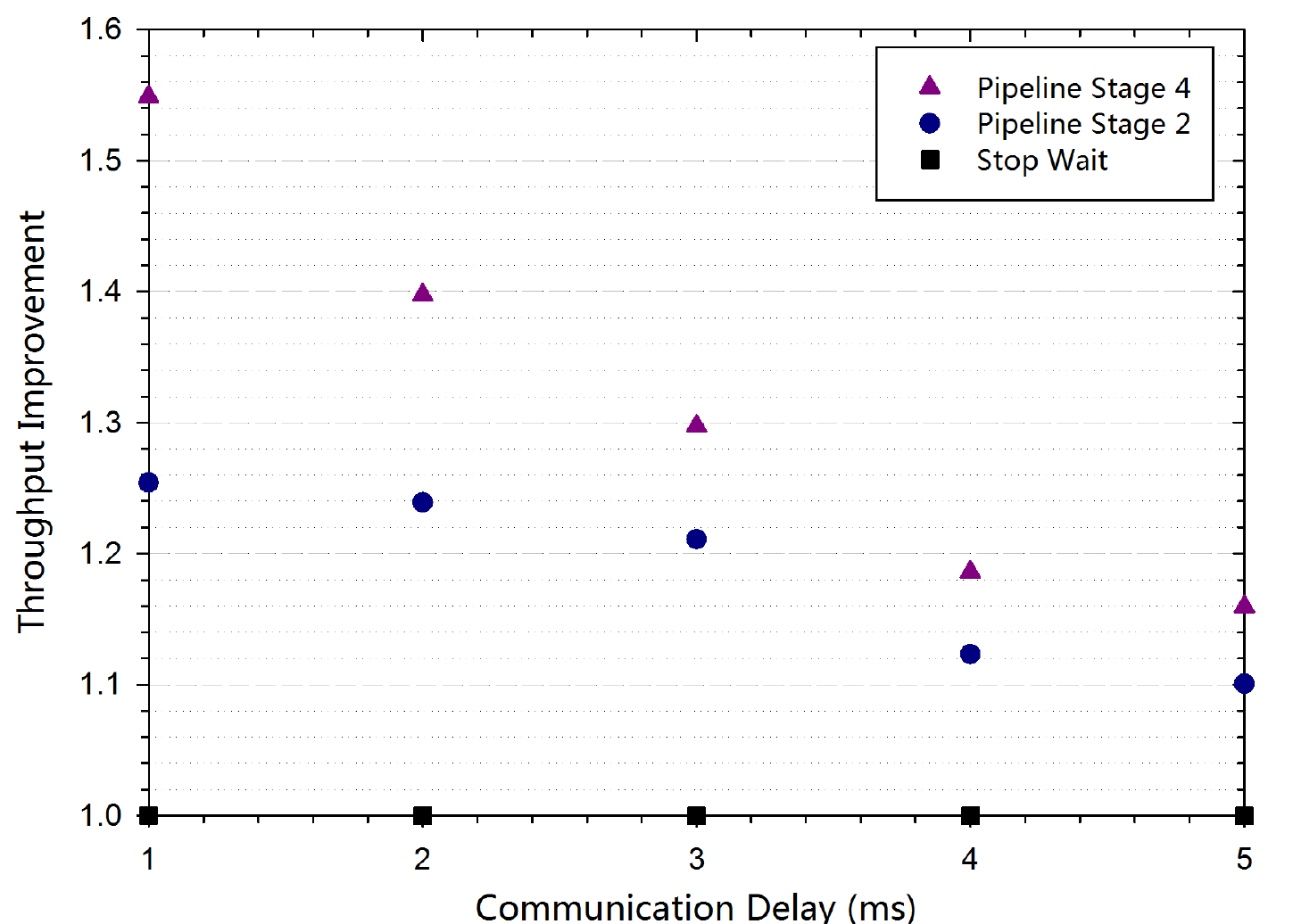}
		\label{fig:4.1}}
	\subfigure[Rates of decline in throughput for the adaptive combination modes]{
		\includegraphics[width=0.41\textwidth]{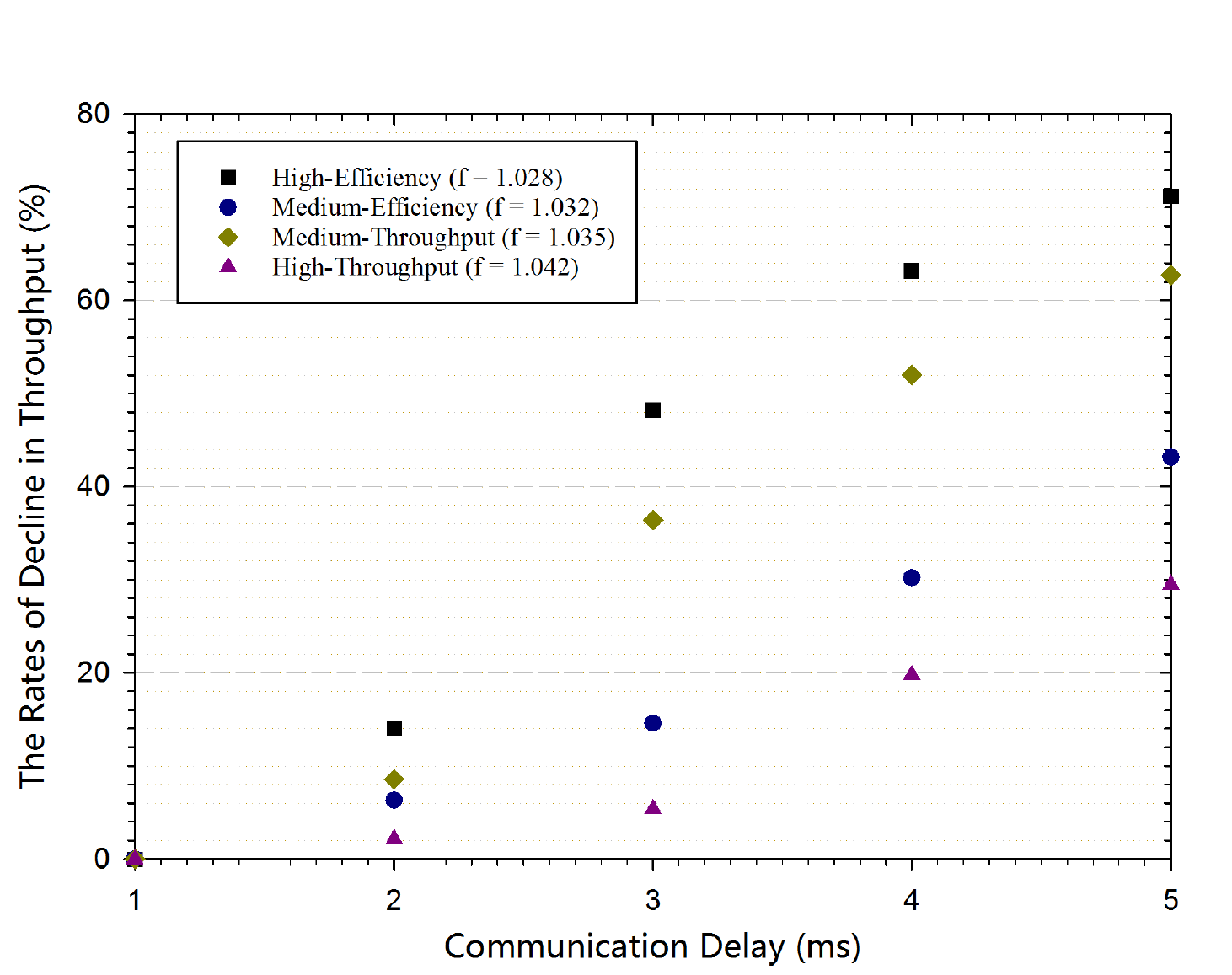}
		\label{fig:4.2}}
	\\
	\subfigure[Effects of backtracking collision detection]{
		\includegraphics[width=0.41\textwidth]{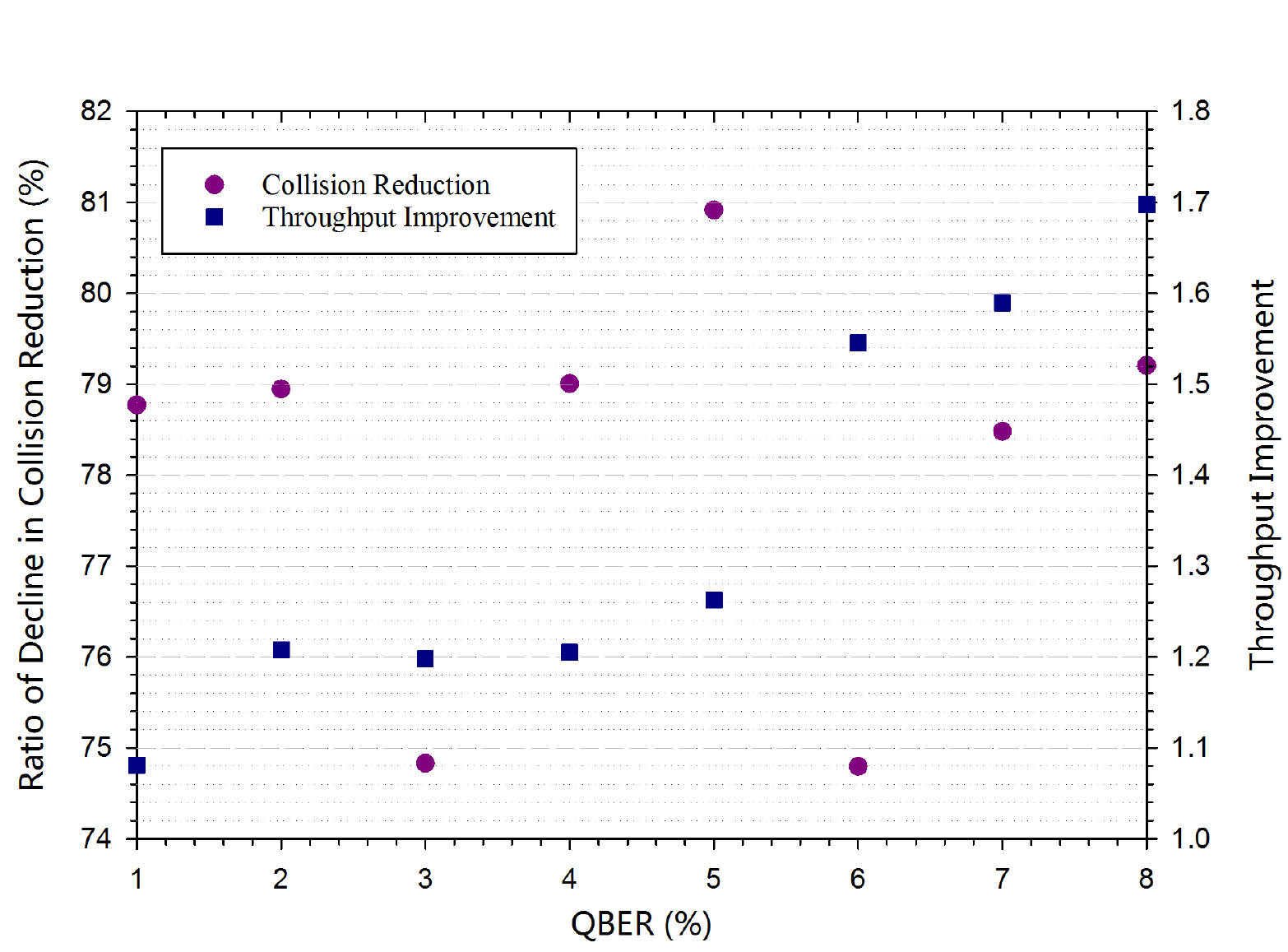}
		\label{fig:4.3}}
	\caption{Experiments for the individual improvements}
	\label{fig:4}
\end{figure}

The experiment results for the individual improvements are shown in Fig. \ref{fig:4}. Fig. \ref{fig:4.1} presents an assessment of the throughput improvement of the multi-pipeline optimization scheme relative to the benchmark of the Stop-Wait scheme in \cite{13_Pedersen_2015}. From it, we can be seen that the percentage improvement in throughput varies from 10\% to 55\%. Though the multi-pipeline scheme offers improvement in the throughput performance, the graph also shows that the effect of each multi-pipeline scheme decreases with the increasing communication delay. This is because the throughput improvement of multi-pipeline scheme mainly comes from resources expanded on computation and communication overhead. However, the contribution of computation overhead to the overall time spent decreases rapidly as the communication delay increases. Consequently, for high communication delays, the communication time spent has been the dominating factor that influences the throughput performance. In this case, the contribution of the multi-pipeline strategy is minimal. 

Next, we analyze the adaptation of different combination modes, which are listed in Table \ref{tab:2}, to communication delays. In the experiment, the throughput with a latency of 1ms is regarded as the baseline and the rates of decline in throughput for different block-length combinations with increasing QBER are plotted in Fig. \ref{fig:4.2}. As seen from the plot, the throughput decreases rapidly with increasing communication delay, especially when the high-efficiency mode is applied. When the communication delay reaches 5ms, the throughput decrease ratio has reaches 70\%. Under the same conditions, the throughput decrease ratio is only 30\% if the high-throughput mode is used. These results indicate that the adaptive combinations of block lengths fit well with the dynamics of communication delays with minimal degradation in efficiency. 

Then the experimental results for backtracking collision detection are presented in Fig. \ref{fig:4.3}. Therefrom we see that the average ratio of decline in collision reduction reaches up to 78\%, which leads to corresponding savings in the communication cost. In addition, the percentage improvement in throughput varies from 8\% to 70\% with the increases in QBER from 1\% to 8\%. This is because the number of blocks and the percentage of backtracking time spent to the total time increase with QBER, thereby leading to a better collision detection effect. Nowadays, a typical DV-QKD system runs under a QBER of less than 3\% for which the percentage improvement in throughput is approximately 20\% \cite{5_Pirandola_2020}.

In addition, since memory usage is a big issue in high performance Cascade implementation, we also compare the actual storage consumption between our work and Ref. \cite{13_Pedersen_2015} in Table \ref{tab:4}. Because we use pipelined storage structure, the memory usage is slightly larger than that in Ref. \cite{13_Pedersen_2015} when QBER equals to 1\%. However, since new storage scheme is applied, the amount of memory used increases significantly slower than Ref. \cite{13_Pedersen_2015}. 

\begin{table}[ht]
\caption{Memory usage}
\label{tab:4}
\centering
\begin{tabular}{cccc}
\hline
Refs. & QBER & Input Size (Mb) & Memory Usage (MB)  \\
\hline
Ours & 1.0\% & 30 & 48  \\ 
Ours & 15.0\% & 30 & 80  \\ 
Ref. \cite{13_Pedersen_2015} & 1.0\% & 30 & 38 \\
Ref. \cite{13_Pedersen_2015} & 15.0\% & 30 & 256 \\
\hline
\end{tabular}
\end{table}

\subsection{Overall Performance Test}
To evaluate the overall performance of our solution in real-world scenarios, we executed it on two computers connected by 50 km dedicated fiber, using 1000Mbps media converters that can convert 1000 Mbps Ethernet connections to a 1000Base-FX fiber connection. In the following experiments, we use four-stage multi-pipelines per processor core. The input buffer size of each pipeline (i.e. thread) is set to 10Mb. Each thread will terminate immediately when its input amount of data reaches 1Gb.

Table \ref{tab:5} presents a comparative analysis of our solution relative to the state-of-the-art CPU \cite{13_Pedersen_2015, 20_Mao_2019}, GPU \cite{9_Dixon_2014} and FPGA \cite{2_Yuan_2018} benchmarks. Among these, Ref. \cite{13_Pedersen_2015} was the previous fastest implementation by using Cascade protocol, while Ref. \cite{20_Mao_2019}, Ref. \cite{9_Dixon_2014} and Ref. \cite{2_Yuan_2018} were the fastest LDPC implementation on CPU, GPU and FPGA platform, respectively. 

\begin{table}[ht]
\caption{The Evaluation Platforms}
\label{tab:5}
\centering
\resizebox{\textwidth}{0.9cm} {
\begin{tabular}{cccccccc}
\hline
     		    & \multicolumn{2}{c}{Ours} & Ref. \cite{13_Pedersen_2015} & \multicolumn{2}{c}{Ref. \cite{20_Mao_2019}} & Ref. \cite{9_Dixon_2014} & Ref. \cite{2_Yuan_2018} \\
\hline
\multirow{2}{*}{Product Collection} & Intel CPU & Intel CPU & Intel CPU & Intel CPU & Intel CPU & NVidia GPU & Intel FPGA \\
                                                         & i7-6700HQ & Atom E3845 & i7 & i7-6700HQ & i9-9900K & M2090 & Altera Stratix V \\
                              Vertical Segment & Mobile & Embedded & Desktop & Mobile & Desktop & Server & - \\
                              Number of Cores & 4 & 4 & 4 & 4 & 8 & 512 & -\\
                                Base Frequency (GHz) & 2.6 & 1.9 & 3.4 & 2.6 & 3.6 & 1.3 & -\\
                      Max Turbo Frequency (GHz) & 3.5 & 1.9 & 3.4 & 3.5 & 5.0 & -  & -\\
                      	    Power Consumption (W) & 45 & 10 & - & 45 & 95 & 225 & -\\
\hline
\end{tabular}
}
\end{table}

\begin{figure}[ht]
	\centering
	\includegraphics[width=0.9\textwidth]{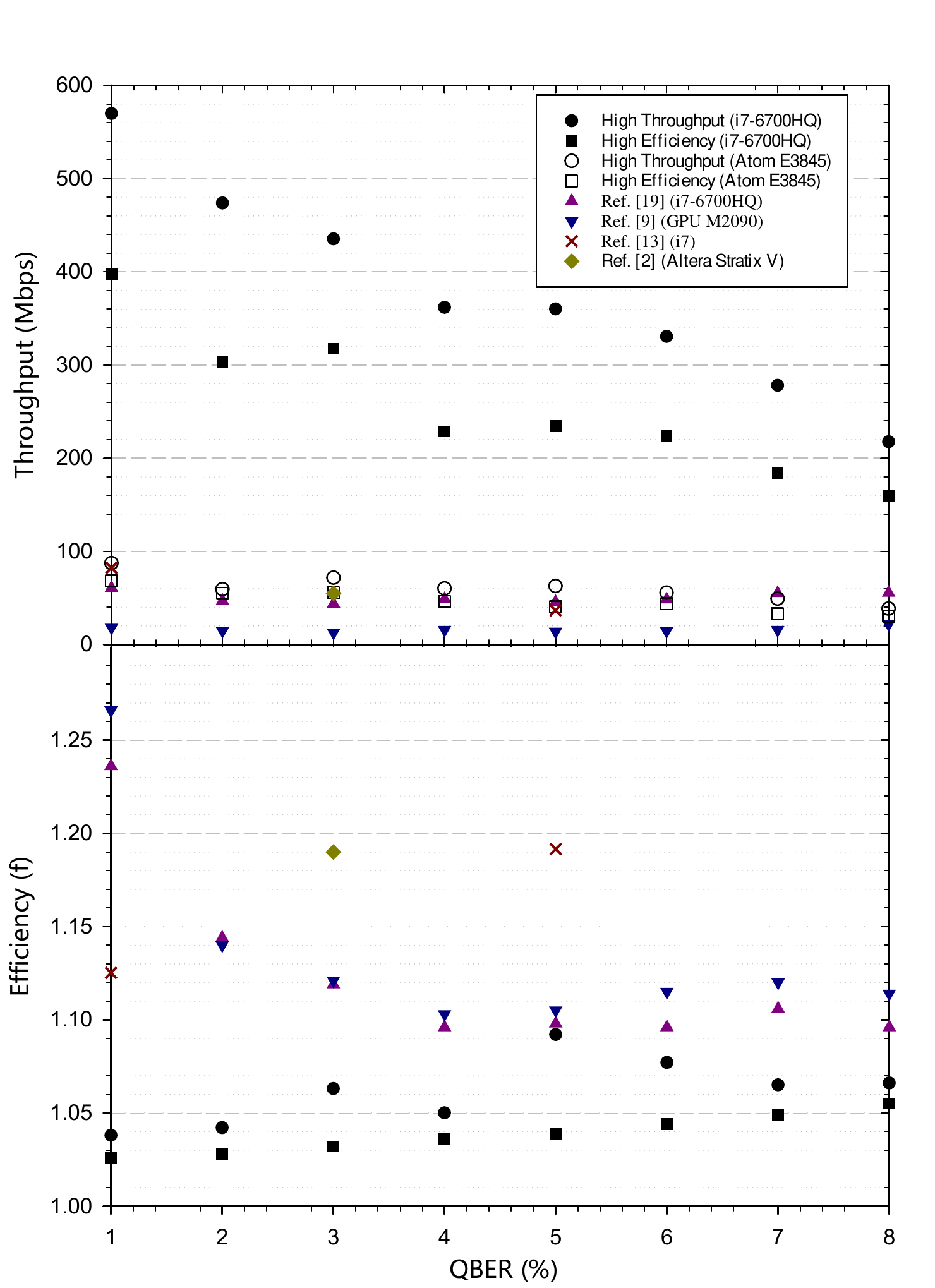}
	\caption{Plot of throughput (upper panel) and efficiency (lower panel) where we note that our efficiencies on i7-6700HQ and Atom E3845 are the same when the same combination mode. }
	\label{fig:5}
\end{figure}

In Fig. \ref{fig:5}, we present a plot of QBER against throughput and efficiency relative to established methods reported in the literature. We note that the throughput of our solution is rather high, but it decreases with increasing QBER. Such a phenomenon is clearly due to the rapidly increasing number of blocks to be processed which subsequently improves the computation cost. Though the throughput correlates inversely with QBER, the throughput and efficiency of our solution on i7-6700HQ surpasses those reported for comparative schemes within the whole range of QBER. To our best knowledge, the previous fastest reconciliation achieved a throughput of 130Mbps with an efficiency of 1.23 when correcting 1\% QBER on two i9-9900K processors \cite{20_Mao_2019}. Compared to this result, a speedup factor of $\times4.4$ is obtained with an efficiency closer to the theoretical limit. Similarly, from the plot in Fig. \ref{fig:5}, we infer that our implementation performs at par with the benchmark methods in terms of level of throughput performance even when the Atom E3845 CPU is applied. We note that Intel Atom E3845 is an embedded platform with rather low power-consumption. Indeed, the performance on Atom E3845 is sufficient for the existing QKD systems. Once the throughput of reconciliation is insufficient, we only need to change a more powerful platform.

To further evaluate the actual performance of our proposed solution, we assessed the performance on real data. By using different operation parameters, we obtained two data sets obtained from a practical QKD system \cite{41_Wang_2018}. Our experiments are validated on Atom E3845 platforms with the high-efficiency combination mode. In addition, we use a pre-defined estimated QBER since the precise QBER is unknown to us. The experiment results are listed in Table \ref{tab:6}. We can see that the throughput of up to 86Mbps with an efficiency of 1.028 was achieved on data set 2. Moreover, even if the estimated QBER is inaccurate, the efficiency degradation is not apparent. This fully validates the good adaptability of our solution to the fluctuating QBER. 

\begin{table}[ht]
\caption{Experiment results on real data sets}
\label{tab:6}
\centering
\begin{tabular}{cccccc}
\hline
Data Set No. & Actual QBER & Estimated QBER & $f$ & $f_{FER}$ & Throughput(Mbps) \\
\hline
\multirow{3}{*}{1} & \multirow{3}{*}{2.0\%} & 1.0\% & 1.033 & 1.037 & 40 \\  
                   &                    & 2.0\% & 1.028 & 1.032 & 57 \\ 
                   &                    & 3.0\% & 1.043 & 1.046 & 76 \\ \hline
\multirow{3}{*}{2} & \multirow{3}{*}{2.6\%} & 1.6\% & 1.024 & 1.028 & 68 \\
                   &                    & 2.6\% & 1.023 & 1.026 & 82 \\
                   &                    & 3.6\% & 1.028 & 1.036 & 86 \\ \hline
\end{tabular}
\end{table}

Encouraged by the outlined performance, it can be surmised that our solution can contribute towards the development of practical QKD systems. It has shown promise in terms of surplus throughput which leads to low performance CPU platforms that are required to reduce overall system costs for most QKD systems (e.g. power consumption, volume, economic cost, etc.). Moreover, once the throughput is adequate to satisfy the processing demand, the secure key rate is determined principally by the reconciliation efficiency. In such case, a high efficiency performance can help to improve the distilling efficiency of the secure keys, thus improving the communication distance and the final secure key rate of a practical QKD system. This performance reinforces our conclusion that the proposed solution is applicable to nearly all discrete-variable (DV) QKD systems.

\section{Conclusions}
In this study, we first proposed a framework to improve the throughput of Cascade reconciliation. The framework increases utilization of computation resources and effectively deals with communication delay. Furthermore, we proposed an integrated-optimization-oriented implementation scheme that targets optimizations of core parameters, storage scheme and backtracking process. Execution of our high-efficiency mode over a 50km fiber, which yielded a throughput of 398Mbps and 318Mbps with the efficiency of 1.026 and 1.032 as well as 1\% and 3\% corrections of QBER, respectively. These outcomes indicate the utility of our proposed solution across nearly all DV-QKD systems.

In ongoing and future work, we are exploring improvements to the proposed solution in two directions. First in terms of efficiency, we are considering the application of a backtracking list to backtrack to the sub-block in the pass of three or later stages of the process. This will facilitate better use of known bits and known parities that may enhance efficiency. Second, in terms of throughput, we are exploring a combination of binary search and hamming error correction as a way to efficiently reduce the number of communication rounds since only one communication round is needed in hamming error correction. However, hamming error correction may introduce new error bit when the block contain three or more errors. Therefore, the focus of the improvements to the scheme should consider designs that help to avoid such unexpected issues.

\backmatter
\bmhead{Acknowledgments}
 Special thanks goes to Mr. Yi Luo for the helpful discussion.

\section*{Declarations}
\begin{itemize}
\item Funding
This work is supported by the National Natural Science Foundation of China (Grant Number: 62071151, 61301099).

\item Conflict of interest/Competing interests
The authors declare no conflicts of interest.

\item Availability of data and materials
Data underlying the results presented in this paper are not publicly available
at this time but may be obtained from the authors upon reasonable request.

\item Code availability 
The code in this paper is not publicly available at this time but may be obtained from the authors upon reasonable request.

\item Authors' contributions
Hao-Kun Mao and Qiong-Li contributed to the initial ideas. Abdullah M. Iliyasu refine the ideas. Peng-Lei Hao and Bassem Abd-El-Atty wrote the source code and completed the simulations. All authors contributed to writing the manuscript.
\end{itemize}




\begin{thebibliography}{37}
\ifx \bisbn   \undefined \def \bisbn  #1{ISBN #1}\fi
\ifx \binits  \undefined \def \binits#1{#1}\fi
\ifx \bauthor  \undefined \def \bauthor#1{#1}\fi
\ifx \batitle  \undefined \def \batitle#1{#1}\fi
\ifx \bjtitle  \undefined \def \bjtitle#1{#1}\fi
\ifx \bvolume  \undefined \def \bvolume#1{\textbf{#1}}\fi
\ifx \byear  \undefined \def \byear#1{#1}\fi
\ifx \bissue  \undefined \def \bissue#1{#1}\fi
\ifx \bfpage  \undefined \def \bfpage#1{#1}\fi
\ifx \blpage  \undefined \def \blpage #1{#1}\fi
\ifx \burl  \undefined \def \burl#1{\textsf{#1}}\fi
\ifx \doiurl  \undefined \def \doiurl#1{\url{https://doi.org/#1}}\fi
\ifx \betal  \undefined \def \betal{\textit{et al.}}\fi
\ifx \binstitute  \undefined \def \binstitute#1{#1}\fi
\ifx \binstitutionaled  \undefined \def \binstitutionaled#1{#1}\fi
\ifx \bctitle  \undefined \def \bctitle#1{#1}\fi
\ifx \beditor  \undefined \def \beditor#1{#1}\fi
\ifx \bpublisher  \undefined \def \bpublisher#1{#1}\fi
\ifx \bbtitle  \undefined \def \bbtitle#1{#1}\fi
\ifx \bedition  \undefined \def \bedition#1{#1}\fi
\ifx \bseriesno  \undefined \def \bseriesno#1{#1}\fi
\ifx \blocation  \undefined \def \blocation#1{#1}\fi
\ifx \bsertitle  \undefined \def \bsertitle#1{#1}\fi
\ifx \bsnm \undefined \def \bsnm#1{#1}\fi
\ifx \bsuffix \undefined \def \bsuffix#1{#1}\fi
\ifx \bparticle \undefined \def \bparticle#1{#1}\fi
\ifx \barticle \undefined \def \barticle#1{#1}\fi
\bibcommenthead
\ifx \bconfdate \undefined \def \bconfdate #1{#1}\fi
\ifx \botherref \undefined \def \botherref #1{#1}\fi
\ifx \url \undefined \def \url#1{\textsf{#1}}\fi
\ifx \bchapter \undefined \def \bchapter#1{#1}\fi
\ifx \bbook \undefined \def \bbook#1{#1}\fi
\ifx \bcomment \undefined \def \bcomment#1{#1}\fi
\ifx \oauthor \undefined \def \oauthor#1{#1}\fi
\ifx \citeauthoryear \undefined \def \citeauthoryear#1{#1}\fi
\ifx \endbibitem  \undefined \def \endbibitem {}\fi
\ifx \bconflocation  \undefined \def \bconflocation#1{#1}\fi
\ifx \arxivurl  \undefined \def \arxivurl#1{\textsf{#1}}\fi
\csname PreBibitemsHook\endcsname

\bibitem{1_Bennett_2014}
\begin{barticle}
\bauthor{\bsnm{Bennett}, \binits{C.H.}},
\bauthor{\bsnm{Brassard}, \binits{G.}}:
\batitle{Quantum cryptography: Public key distribution and coin tossing}.
\bjtitle{Theoretical Computer Science}
\bvolume{560},
\bfpage{7}--\blpage{11}
(\byear{2014})
\end{barticle}
\endbibitem

\bibitem{2_Yuan_2018}
\begin{barticle}
\bauthor{\bsnm{Yuan}, \binits{Z.}},
\bauthor{\bsnm{Murakami}, \binits{A.}},
\bauthor{\bsnm{Kujiraoka}, \binits{M.}},
\bauthor{\bsnm{Lucamarini}, \binits{M.}},
\bauthor{\bsnm{Tanizawa}, \binits{Y.}},
\bauthor{\bsnm{Sato}, \binits{H.}},
\bauthor{\bsnm{Shields}, \binits{A.J.}},
\bauthor{\bsnm{Plews}, \binits{A.}},
\bauthor{\bsnm{Takahashi}, \binits{R.}},
\bauthor{\bsnm{Doi}, \binits{K.}},
\bauthor{\bsnm{Tam}, \binits{W.}},
\bauthor{\bsnm{Sharpe}, \binits{A.W.}},
\bauthor{\bsnm{Dixon}, \binits{A.R.}},
\bauthor{\bsnm{Lavelle}, \binits{E.}},
\bauthor{\bsnm{Dynes}, \binits{J.F.}}:
\batitle{10-mb/s quantum key distribution}.
\bjtitle{Journal of Lightwave Technology}
\bvolume{36}(\bissue{16}),
\bfpage{3427}--\blpage{3433}
(\byear{2018})
\end{barticle}
\endbibitem

\bibitem{3_Duplinskiy_2018}
\begin{barticle}
\bauthor{\bsnm{Duplinskiy}, \binits{A.V.}},
\bauthor{\bsnm{Kiktenko}, \binits{E.O.}},
\bauthor{\bsnm{Pozhar}, \binits{N.O.}},
\bauthor{\bsnm{Anufriev}, \binits{M.N.}},
\bauthor{\bsnm{Ermakov}, \binits{R.P.}},
\bauthor{\bsnm{Kotov}, \binits{A.I.}},
\bauthor{\bsnm{Brodskiy}, \binits{A.V.}},
\bauthor{\bsnm{Yunusov}, \binits{R.R.}},
\bauthor{\bsnm{Kurochkin}, \binits{V.L.}},
\bauthor{\bsnm{Fedorov}, \binits{A.K.}},
\bauthor{\bsnm{Kurochkin}, \binits{Y.V.}}:
\batitle{Quantum-secured data transmission in urban fiber-optics communication
  lines}.
\bjtitle{Journal of Russian Laser Research}
\bvolume{39}(\bissue{2}),
\bfpage{113}--\blpage{119}
(\byear{2018})
\end{barticle}
\endbibitem

\bibitem{4_Boaron_2018}
\begin{barticle}
\bauthor{\bsnm{Boaron}, \binits{A.}},
\bauthor{\bsnm{Korzh}, \binits{B.}},
\bauthor{\bsnm{Houlmann}, \binits{R.}},
\bauthor{\bsnm{Boso}, \binits{G.}},
\bauthor{\bsnm{Rusca}, \binits{D.}},
\bauthor{\bsnm{Gray}, \binits{S.}},
\bauthor{\bsnm{Li}, \binits{M.-J.}},
\bauthor{\bsnm{Nolan}, \binits{D.}},
\bauthor{\bsnm{Martin}, \binits{A.}},
\bauthor{\bsnm{Zbinden}, \binits{H.}}:
\batitle{Simple 2.5{GHz} time-bin quantum key distribution}.
\bjtitle{Applied Physics Letters}
\bvolume{112}(\bissue{17}),
\bfpage{171108}
(\byear{2018})
\end{barticle}
\endbibitem

\bibitem{5_Pirandola_2020}
\begin{barticle}
\bauthor{\bsnm{Pirandola}, \binits{S.}},
\bauthor{\bsnm{Andersen}, \binits{U.L.}},
\bauthor{\bsnm{Banchi}, \binits{L.}},
\bauthor{\bsnm{Berta}, \binits{M.}},
\bauthor{\bsnm{Bunandar}, \binits{D.}},
\bauthor{\bsnm{Colbeck}, \binits{R.}},
\bauthor{\bsnm{Englund}, \binits{D.}},
\bauthor{\bsnm{Gehring}, \binits{T.}},
\bauthor{\bsnm{Lupo}, \binits{C.}},
\bauthor{\bsnm{Ottaviani}, \binits{C.}},
\bauthor{\bsnm{Pereira}, \binits{J.L.}},
\bauthor{\bsnm{Razavi}, \binits{M.}},
\bauthor{\bsnm{Shaari}, \binits{J.S.}},
\bauthor{\bsnm{Tomamichel}, \binits{M.}},
\bauthor{\bsnm{Usenko}, \binits{V.C.}},
\bauthor{\bsnm{Vallone}, \binits{G.}},
\bauthor{\bsnm{Villoresi}, \binits{P.}},
\bauthor{\bsnm{Wallden}, \binits{P.}}:
\batitle{Advances in quantum cryptography}.
\bjtitle{Advances in Optics and Photonics}
\bvolume{12}(\bissue{4}),
\bfpage{1012}
(\byear{2020})
\end{barticle}
\endbibitem

\bibitem{6_Lucamarini_2018}
\begin{barticle}
\bauthor{\bsnm{Lucamarini}, \binits{M.}},
\bauthor{\bsnm{Yuan}, \binits{Z.L.}},
\bauthor{\bsnm{Dynes}, \binits{J.F.}},
\bauthor{\bsnm{Shields}, \binits{A.J.}}:
\batitle{Overcoming the rate{\textendash}distance limit of quantum key
  distribution without quantum repeaters}.
\bjtitle{Nature}
\bvolume{557}(\bissue{7705}),
\bfpage{400}--\blpage{403}
(\byear{2018})
\end{barticle}
\endbibitem

\bibitem{7_Yin_2016}
\begin{botherref}
\oauthor{\bsnm{Yin}, \binits{H.-L.}},
\oauthor{\bsnm{Chen}, \binits{T.-Y.}},
\oauthor{\bsnm{Yu}, \binits{Z.-W.}},
\oauthor{\bsnm{Liu}, \binits{H.}},
\oauthor{\bsnm{You}, \binits{L.-X.}},
\oauthor{\bsnm{Zhou}, \binits{Y.-H.}},
\oauthor{\bsnm{Chen}, \binits{S.-J.}},
\oauthor{\bsnm{Mao}, \binits{Y.}},
\oauthor{\bsnm{Huang}, \binits{M.-Q.}},
\oauthor{\bsnm{Zhang}, \binits{W.-J.}},
\oauthor{\bsnm{Chen}, \binits{H.}},
\oauthor{\bsnm{Li}, \binits{M.J.}},
\oauthor{\bsnm{Nolan}, \binits{D.}},
\oauthor{\bsnm{Zhou}, \binits{F.}},
\oauthor{\bsnm{Jiang}, \binits{X.}},
\oauthor{\bsnm{Wang}, \binits{Z.}},
\oauthor{\bsnm{Zhang}, \binits{Q.}},
\oauthor{\bsnm{Wang}, \binits{X.-B.}},
\oauthor{\bsnm{Pan}, \binits{J.-W.}}:
Measurement-device-independent quantum key distribution over a 404~km optical
  fiber.
Physical Review Letters
\textbf{117}(19)
(2016)
\end{botherref}
\endbibitem

\bibitem{8_Zhang_2020}
\begin{botherref}
\oauthor{\bsnm{Zhang}, \binits{Y.}},
\oauthor{\bsnm{Chen}, \binits{Z.}},
\oauthor{\bsnm{Pirandola}, \binits{S.}},
\oauthor{\bsnm{Wang}, \binits{X.}},
\oauthor{\bsnm{Zhou}, \binits{C.}},
\oauthor{\bsnm{Chu}, \binits{B.}},
\oauthor{\bsnm{Zhao}, \binits{Y.}},
\oauthor{\bsnm{Xu}, \binits{B.}},
\oauthor{\bsnm{Yu}, \binits{S.}},
\oauthor{\bsnm{Guo}, \binits{H.}}:
Long-distance continuous-variable quantum key distribution over 202.81~km of
  fiber.
Physical Review Letters
\textbf{125}(1)
(2020)
\end{botherref}
\endbibitem

\bibitem{9_Dixon_2014}
\begin{botherref}
\oauthor{\bsnm{Dixon}, \binits{A.R.}},
\oauthor{\bsnm{Sato}, \binits{H.}}:
High speed and adaptable error correction for megabit/s rate quantum key
  distribution.
Scientific Reports
\textbf{4}(1)
(2014)
\end{botherref}
\endbibitem

\bibitem{10_Wang_2018}
\begin{botherref}
\oauthor{\bsnm{Wang}, \binits{X.}},
\oauthor{\bsnm{Zhang}, \binits{Y.}},
\oauthor{\bsnm{Yu}, \binits{S.}},
\oauthor{\bsnm{Guo}, \binits{H.}}:
High speed error correction for continuous-variable quantum key distribution
  with multi-edge type {LDPC} code.
Scientific Reports
\textbf{8}(1)
(2018)
\end{botherref}
\endbibitem

\bibitem{11_Gao_2019}
\begin{barticle}
\bauthor{\bsnm{Gao}, \binits{C.}},
\bauthor{\bsnm{Jiang}, \binits{D.}},
\bauthor{\bsnm{Guo}, \binits{Y.}},
\bauthor{\bsnm{Chen}, \binits{L.}}:
\batitle{Multi-matrix error estimation and reconciliation for quantum key
  distribution}.
\bjtitle{Optics Express}
\bvolume{27}(\bissue{10}),
\bfpage{14545}
(\byear{2019})
\end{barticle}
\endbibitem

\bibitem{12_abd2020controlled}
\begin{barticle}
\bauthor{\bsnm{Abd~EL-Latif}, \binits{A.A.}},
\bauthor{\bsnm{Abd-El-Atty}, \binits{B.}},
\bauthor{\bsnm{Abou-Nassar}, \binits{E.M.}},
\bauthor{\bsnm{Venegas-Andraca}, \binits{S.E.}}:
\batitle{Controlled alternate quantum walks based privacy preserving healthcare
  images in internet of things}.
\bjtitle{Optics \& Laser Technology}
\bvolume{124},
\bfpage{105942}
(\byear{2020})
\end{barticle}
\endbibitem

\bibitem{13_Pedersen_2015}
\begin{barticle}
\bauthor{\bsnm{Pedersen}, \binits{T.B.}},
\bauthor{\bsnm{Toyran}, \binits{M.}}:
\batitle{High performance information reconciliation for qkd with cascade}.
\bjtitle{Quantum Info. Comput.}
\bvolume{15}(\bissue{5–-6}),
\bfpage{419}--\blpage{434}
(\byear{2015})
\end{barticle}
\endbibitem

\bibitem{14_abd2020quantum}
\begin{barticle}
\bauthor{\bsnm{Abd~el-Latif}, \binits{A.A.}},
\bauthor{\bsnm{Abd-el-Atty}, \binits{B.}},
\bauthor{\bsnm{Amin}, \binits{M.}},
\bauthor{\bsnm{Iliyasu}, \binits{A.M.}}:
\batitle{Quantum-inspired cascaded discrete-time quantum walks with induced
  chaotic dynamics and cryptographic applications}.
\bjtitle{Scientific reports}
\bvolume{10}(\bissue{1}),
\bfpage{1}--\blpage{16}
(\byear{2020})
\end{barticle}
\endbibitem

\bibitem{15_abd2019efficient}
\begin{barticle}
\bauthor{\bsnm{Abd~EL-Latif}, \binits{A.A.}},
\bauthor{\bsnm{Abd-El-Atty}, \binits{B.}},
\bauthor{\bsnm{Venegas-Andraca}, \binits{S.E.}},
\bauthor{\bsnm{Mazurczyk}, \binits{W.}}:
\batitle{Efficient quantum-based security protocols for information sharing and
  data protection in 5g networks}.
\bjtitle{Future Generation Computer Systems}
\bvolume{100},
\bfpage{893}--\blpage{906}
(\byear{2019})
\end{barticle}
\endbibitem

\bibitem{16_abd2020secure}
\begin{barticle}
\bauthor{\bsnm{Abd~El-Latif}, \binits{A.A.}},
\bauthor{\bsnm{Abd-El-Atty}, \binits{B.}},
\bauthor{\bsnm{Mazurczyk}, \binits{W.}},
\bauthor{\bsnm{Fung}, \binits{C.}},
\bauthor{\bsnm{Venegas-Andraca}, \binits{S.E.}}:
\batitle{Secure data encryption based on quantum walks for 5g internet of
  things scenario}.
\bjtitle{IEEE Transactions on Network and Service Management}
\bvolume{17}(\bissue{1}),
\bfpage{118}--\blpage{131}
(\byear{2020})
\end{barticle}
\endbibitem

\bibitem{17_Li_2014}
\begin{barticle}
\bauthor{\bsnm{Qiong}, \binits{L.}},
\bauthor{\bsnm{Dan}, \binits{L.}},
\bauthor{\bsnm{Haokun}, \binits{M.}},
\bauthor{\bsnm{Xiamu}, \binits{N.}},
\bauthor{\bsnm{Tian}, \binits{L.}},
\bauthor{\bsnm{Hong}, \binits{G.}}:
\batitle{Study on error reconciliation in quantum key distribution}.
\bjtitle{Quantum Info. Comput.}
\bvolume{14}(\bissue{13--14}),
\bfpage{1117}--\blpage{1135}
(\byear{2014})
\end{barticle}
\endbibitem

\bibitem{18_Brassard_1993}
\begin{botherref}
\oauthor{\bsnm{Brassard}, \binits{G.}},
\oauthor{\bsnm{Salvail}, \binits{L.}}:
Secret-key reconciliation by public discussion.
In: Advances in Cryptology {\textemdash} {EUROCRYPT} '93,
pp. 410--423.
Springer
\end{botherref}
\endbibitem

\bibitem{20_Mao_2019}
\begin{botherref}
\oauthor{\bsnm{Mao}, \binits{H.}},
\oauthor{\bsnm{Li}, \binits{Q.}},
\oauthor{\bsnm{Han}, \binits{Q.}},
\oauthor{\bsnm{Guo}, \binits{H.}}:
High-throughput and low-cost {LDPC} reconciliation for quantum key
  distribution.
Quantum Information Processing
\textbf{18}(7)
(2019)
\end{botherref}
\endbibitem

\bibitem{21_Jesus_2015}
\begin{barticle}
\bauthor{\bsnm{Martinez-Mateo}, \binits{J.}},
\bauthor{\bsnm{Pacher}, \binits{C.}},
\bauthor{\bsnm{Peev}, \binits{M.}},
\bauthor{\bsnm{Ciurana}, \binits{A.}},
\bauthor{\bsnm{Martin}, \binits{V.}}:
\batitle{Demystifying the information reconciliation protocol cascade}.
\bjtitle{Quantum Info. Comput.}
\bvolume{15}(\bissue{5–-6}),
\bfpage{453}--\blpage{477}
(\byear{2015})
\end{barticle}
\endbibitem

\bibitem{22_Yan_2008}
\begin{bchapter}
\bauthor{\bsnm{Yan}, \binits{H.}},
\bauthor{\bsnm{Ren}, \binits{T.}},
\bauthor{\bsnm{Peng}, \binits{X.}},
\bauthor{\bsnm{Lin}, \binits{X.}},
\bauthor{\bsnm{Jiang}, \binits{W.}},
\bauthor{\bsnm{Liu}, \binits{T.}},
\bauthor{\bsnm{Guo}, \binits{H.}}:
\bctitle{Information reconciliation protocol in quantum key distribution
  system}.
In: \bbtitle{2008 Fourth International Conference on Natural Computation}
(\byear{2008})
\end{bchapter}
\endbibitem

\bibitem{23_Pacher_2015}
\begin{bchapter}
\bauthor{\bsnm{Pacher}, \binits{C.}},
\bauthor{\bsnm{Grabenweger}, \binits{P.}},
\bauthor{\bsnm{Martinez-Mateo}, \binits{J.}},
\bauthor{\bsnm{Martin}, \binits{V.}}:
\bctitle{An information reconciliation protocol for secret-key agreement with
  small leakage}.
In: \bbtitle{2015 {IEEE} International Symposium on Information Theory
  ({ISIT})}
(\byear{2015})
\end{bchapter}
\endbibitem

\bibitem{25_David_2011}
\begin{barticle}
\bauthor{\bsnm{Elkouss}, \binits{D.}},
\bauthor{\bsnm{Martinez-mateo}, \binits{J.}},
\bauthor{\bsnm{Martin}, \binits{V.}}:
\batitle{Information reconciliation for quantum key distribution}.
\bjtitle{Quantum Info. Comput.}
\bvolume{11}(\bissue{3}),
\bfpage{226}--\blpage{238}
(\byear{2011})
\end{barticle}
\endbibitem

\bibitem{26_Jesus_2012}
\begin{barticle}
\bauthor{\bsnm{Martinez-Mateo}, \binits{J.}},
\bauthor{\bsnm{Elkouss}, \binits{D.}},
\bauthor{\bsnm{Martin}, \binits{V.}}:
\batitle{Blind reconciliation}.
\bjtitle{Quantum Info. Comput.}
\bvolume{12}(\bissue{9–-10}),
\bfpage{791}--\blpage{812}
(\byear{2012})
\end{barticle}
\endbibitem

\bibitem{27_Kiktenko_2017}
\begin{botherref}
\oauthor{\bsnm{Kiktenko}, \binits{E.O.}},
\oauthor{\bsnm{Trushechkin}, \binits{A.S.}},
\oauthor{\bsnm{Lim}, \binits{C.C.W.}},
\oauthor{\bsnm{Kurochkin}, \binits{Y.V.}},
\oauthor{\bsnm{Fedorov}, \binits{A.K.}}:
Symmetric blind information reconciliation for quantum key distribution.
Physical Review Applied
\textbf{8}(4)
(2017)
\end{botherref}
\endbibitem

\bibitem{29_Li_2020}
\begin{botherref}
\oauthor{\bsnm{Li}, \binits{Y.}},
\oauthor{\bsnm{Zhang}, \binits{X.}},
\oauthor{\bsnm{Li}, \binits{Y.}},
\oauthor{\bsnm{Xu}, \binits{B.}},
\oauthor{\bsnm{Ma}, \binits{L.}},
\oauthor{\bsnm{Yang}, \binits{J.}},
\oauthor{\bsnm{Huang}, \binits{W.}}:
High-throughput {GPU} layered decoder of quasi-cyclic multi-edge type low
  density parity check codes in continuous-variable quantum key distribution
  systems.
Scientific Reports
\textbf{10}(1)
(2020)
\end{botherref}
\endbibitem

\bibitem{30_Yang_2020}
\begin{barticle}
\bauthor{\bsnm{Yang}, \binits{S.-S.}},
\bauthor{\bsnm{Lu}, \binits{Z.-G.}},
\bauthor{\bsnm{Li}, \binits{Y.-M.}}:
\batitle{High-speed post-processing in continuous-variable quantum key
  distribution based on {FPGA} implementation}.
\bjtitle{Journal of Lightwave Technology}
\bvolume{38}(\bissue{15}),
\bfpage{3935}--\blpage{3941}
(\byear{2020})
\end{barticle}
\endbibitem

\bibitem{31_Paul_2014}
\begin{barticle}
\bauthor{\bsnm{Jouguet}, \binits{P.}},
\bauthor{\bsnm{Kunz-Jacques}, \binits{S.}}:
\batitle{High performance error correction for quantum key distribution using
  polar codes}.
\bjtitle{Quantum Info. Comput.}
\bvolume{14}(\bissue{3–-4}),
\bfpage{329}--\blpage{338}
(\byear{2014})
\end{barticle}
\endbibitem

\bibitem{32_Yan_2018}
\begin{barticle}
\bauthor{\bsnm{Yan}, \binits{S.}},
\bauthor{\bsnm{Wang}, \binits{J.}},
\bauthor{\bsnm{Fang}, \binits{J.}},
\bauthor{\bsnm{Jiang}, \binits{L.}},
\bauthor{\bsnm{Wang}, \binits{X.}}:
\batitle{An improved polar codes-based key reconciliation for practical quantum
  key distribution}.
\bjtitle{Chinese Journal of Electronics}
\bvolume{27}(\bissue{2}),
\bfpage{250}--\blpage{255}
(\byear{2018})
\end{barticle}
\endbibitem

\bibitem{33_Kiktenko_2020}
\begin{botherref}
\oauthor{\bsnm{Kiktenko}, \binits{E.O.}},
\oauthor{\bsnm{Malyshev}, \binits{A.O.}},
\oauthor{\bsnm{Fedorov}, \binits{A.K.}}:
Blind information reconciliation with polar codes for quantum key distribution.
{IEEE} Communications Letters,
1--1
(2020)
\end{botherref}
\endbibitem

\bibitem{34_Fung_2010}
\begin{botherref}
\oauthor{\bsnm{Fung}, \binits{C.-H.F.}},
\oauthor{\bsnm{Ma}, \binits{X.}},
\oauthor{\bsnm{Chau}, \binits{H.F.}}:
Practical issues in quantum-key-distribution postprocessing.
Physical Review A
\textbf{81}(1)
(2010)
\end{botherref}
\endbibitem

\bibitem{35_Yan_2020}
\begin{barticle}
\bauthor{\bsnm{Yan}, \binits{B.}},
\bauthor{\bsnm{Li}, \binits{Q.}},
\bauthor{\bsnm{Mao}, \binits{H.}},
\bauthor{\bsnm{Xue}, \binits{X.}}:
\batitle{High-speed privacy amplification scheme using {GMP} in quantum key
  distribution}.
\bjtitle{{IEEE} Photonics Journal}
\bvolume{12}(\bissue{3}),
\bfpage{1}--\blpage{13}
(\byear{2020})
\end{barticle}
\endbibitem

\bibitem{36_toyran2017new}
\begin{bchapter}
\bauthor{\bsnm{Toyran}, \binits{M.}},
\bauthor{\bsnm{Toyran}, \binits{M.}},
\bauthor{\bsnm{{\"O}zt{\"u}rk}, \binits{S.}}:
\bctitle{New approaches to increase efficiency of cascade information
  reconciliation protocol}.
In: \bbtitle{7th International Conference on Quantum Cryptography, Cambridge,
  UK}
(\byear{2017})
\end{bchapter}
\endbibitem

\bibitem{37_Le_Gal_2016}
\begin{barticle}
\bauthor{\bsnm{Gal}, \binits{B.L.}},
\bauthor{\bsnm{Jego}, \binits{C.}}:
\batitle{High-throughput multi-core {LDPC} decoders based on x86 processor}.
\bjtitle{{IEEE} Transactions on Parallel and Distributed Systems}
\bvolume{27}(\bissue{5}),
\bfpage{1373}--\blpage{1386}
(\byear{2016})
\end{barticle}
\endbibitem

\bibitem{38_Levinthal_2009}
\begin{barticle}
\bauthor{\bsnm{Levinthal}, \binits{D.}}:
\batitle{Performance analysis guide for intel core i7 processor and intel xeon
  5500 processors}.
\bjtitle{Intel Performance Analysis Guide}
\bvolume{30},
\bfpage{18}
(\byear{2009})
\end{barticle}
\endbibitem

\bibitem{40_Li_2018}
\begin{bchapter}
\bauthor{\bsnm{Li}, \binits{Q.}},
\bauthor{\bsnm{Yang}, \binits{Z.}},
\bauthor{\bsnm{Mao}, \binits{H.}},
\bauthor{\bsnm{Wang}, \binits{X.}}:
\bctitle{Study on scrambling algorithms of error reconciliation in {QKD}}.
In: \bbtitle{2018 Eighth International Conference on Instrumentation {\&}
  Measurement, Computer, Communication and Control ({IMCCC})}
(\byear{2018})
\end{bchapter}
\endbibitem

\bibitem{41_Wang_2018}
\begin{barticle}
\bauthor{\bsnm{Wang}, \binits{S.}},
\bauthor{\bsnm{Chen}, \binits{W.}},
\bauthor{\bsnm{Yin}, \binits{Z.-Q.}},
\bauthor{\bsnm{He}, \binits{D.-Y.}},
\bauthor{\bsnm{Hui}, \binits{C.}},
\bauthor{\bsnm{Hao}, \binits{P.-L.}},
\bauthor{\bsnm{Fan-Yuan}, \binits{G.-J.}},
\bauthor{\bsnm{Wang}, \binits{C.}},
\bauthor{\bsnm{Zhang}, \binits{L.-J.}},
\bauthor{\bsnm{Kuang}, \binits{J.}},
\bauthor{\bsnm{Liu}, \binits{S.-F.}},
\bauthor{\bsnm{Zhou}, \binits{Z.}},
\bauthor{\bsnm{Wang}, \binits{Y.-G.}},
\bauthor{\bsnm{Guo}, \binits{G.-C.}},
\bauthor{\bsnm{Han}, \binits{Z.-F.}}:
\batitle{Practical gigahertz quantum key distribution robust against channel
  disturbance}.
\bjtitle{Optics Letters}
\bvolume{43}(\bissue{9}),
\bfpage{2030}
(\byear{2018})
\end{barticle}
\endbibitem

\end{thebibliography}


\end{document}